\documentclass[runningheads]{llncs}
\usepackage{url}

\usepackage{venndiagram}
\usepackage{color, soul}
\usepackage[latin1,utf8]{inputenc}
\usepackage[linesnumbered,ruled,vlined]{algorithm2e}
\usepackage{graphicx}
\usepackage{float}
\usepackage{tikz}
\usepackage{caption}
\usetikzlibrary{shapes,backgrounds}
\usetikzlibrary{spy}
\usepackage{amsmath}
\usepackage{amssymb}
\usepackage[justification=centering]{subfig}
\usepackage{mathtools} 

\newcommand*{\field}[1]{\mathbb{#1}}%

\let\oldnl\nl
\newcommand{\nonl}{\renewcommand{\nl}{\let\nl\oldnl}}
\DeclareUnicodeCharacter{00A0}{ }

\begin{document}
\title{Multi-party Private Set Operations with an External Decider}
\titlerunning{Multi-party PSO with an External Decider}

\author{Sara Ramezanian \and Tommi Meskanen \and
Valtteri Niemi}
\authorrunning{S. Ramezanian et al.}
%
\institute{University of Helsinki, Department of Computer Science\\
P.O. Box 68 (Pietari Kalmin katu 5), FI-00014, Helsinki, Finland\\
Helsinki Institute for Information Technology (HIIT)\\
\email{sara.ramezanian@helsinki.fi, tommi.meskanen@helsinki.fi, valtteri.niemi@helsinki.fi}\\ }

\maketitle              
\begin{abstract}
A Private Set Operation (PSO) protocol involves at least two parties with their private input sets. The goal of the protocol is for the parties to learn the output of a set operation, i.e. set intersection, on their input sets, without revealing any information about the items that are not in the output set. Commonly, the outcome of the set operation is revealed to parties and no-one else. However, in many application areas of PSO the result of the set operation should be learned by an external participant whom does not have an input set. We call this participant \textit{the decider}. In this paper, we present new variants of multi-party PSO, where there is a decider who gets the result. All parties expect the decider have a private set. Other parties neither learn this result, nor anything else about this protocol. Moreover, we present a generic solution to the problem of PSO.

\end{abstract}
\section{Introduction}
Private Set Operations (PSO) \cite{kissner2005privacy} such as Private Set Intersection (PSI) \cite{kolesnikov2016efficient}, Private Set Union (PSU) \cite{frikken2007privacy}, and Private Membership Test (PMT) \cite{ramezanian2020private}, are special cases of Secure Multi-party Computation that have been in the interest of many researchers. In particular, the problem of PSI has been studied a lot and many PSI protocols have been proposed \cite{pinkas2018scalable}. 

Electronic voting \cite{mursi2013development}, botnet detection between different ISPs \cite{nagaraja2010botgrep}, and genomic applications \cite{erlich2014routes} are a few real-life examples of set operations where the sets are about sensitive data. Also, there is a growing need for privacy-preserving data mining, e.g., in the context of 5G networks. Therefore, a comprehensive study of the factors that affect the feasibility and efficiency of a PSO protocol is needed, and a general but still feasible solution to any multi-part PSO problem is worth seeking for.

Multi-party PSO is a protocol with several parties, each holding a private set, who want to perform a function on their input sets. At the end of the protocol, usually all parties learn the outcome of this function, and nothing else. However, in many of the realistic PSO scenarios, instead of parties themselves learning the outcome, there is a special party who does not have an input set but who needs to learn the result of the PSO. We call this special party \textit{an external decider}, or for short, \textit{a decider}. In Section 2 we present several examples of real life scenarios where the result of PSO is obtained only by the decider, and no-one else.

The contributions of this work are as follows:
\vspace{-0.1cm}
\begin{itemize}
\item We classify the problem of PSO with different criteria.
\item We present a comprehensive study of PSO problem with an external decider. To the best of our knowledge, this variant of PSO has been studied only in special cases, such as secure electronic voting, but not in the general case.
\item We specifically study the case where the set elements are chosen from a universe of limited size. We present a general solution to any PSO problem with external decider and with limited universe.
\item We present another general solution to any PSO problem with external decider, where the universe is not limited. This protocol solves the emptiness and cardinality of the output set.
\item We assume that all parties are semi-honest, but we also present a modification that provides protection in the presence of an individual malicious party.
\item Finally, we implement our protocols and compare the efficiency of our protocols against the state of the art in PSO problems.
\end{itemize} 

Next we briefly explain the necessary concepts that are required to understand the rest of this paper.\\
\textbf{Homomorphic encryption (HE) schemes} allow computations to be carried out using encrypted values, without the need to decrypt them first. In our paper, we are interested in {\it additively} homomorphic encryption such as Paillier cryptosystem \cite{paillier1999public}. If two cipher texts $c_1=enc(m_1)$ and $c_2=enc(m_2)$, are generated by using an additively homomorphic encryption scheme, then the product (or result of some other operation) of these two ciphertexts is decrypted to the sum of the two plaintexts: $dec(c_1c_2)=m_1+m_2$.\\
\textbf{Keyed hash function}, such as HMAC \cite{krawczyk1997hmac}, is a cryptographic hash function with a secret key that is utilized to create fingerprints of a message. The key is only known to the trusted parties. A keyed hash function is a collision-resistant one-way function. \cite{turner2008keyed}.

\section{Motivational Examples of PSO with the Decider}
In this section, we present several examples of scenarios where most parties provide only input data while the result of the set operation goes to an external decider. In each scenario preserving the privacy is important.
\begin{itemize}
\item \textbf{Example-1: }The World Health Organization (WHO) wants to know the number of people in the world who are HIV-positive. The result is the number of all patients in all the hospitals without revealing names. Health data is privacy-sensitive, and therefore, a secure PSO is needed to perform this task. This is an example where the decider (WHO) learns the cardinality of the union of the input sets (patients in hospitals).
\item \textbf{Example-2: } Several proposals are on the table for a board meeting of an organization. Every board member would choose which proposals are acceptable for them. However, they are not willing to reveal their choices to others. Therefore, a PSO protocol is used and the secretary (the decider) computes what proposals are acceptable for everyone. 
\item \textbf{Example-3: }A traffic office has installed surveillance cameras in a city and each of them is collecting plate numbers of the passing cars. The office wants to collect all kind of statistics from the traffic. For example, how many different cars were observed during a day, or how many of those cars that were seen at either point $A$ or point $B$ in the morning were also seen at one of these points in the afternoon. Of course, technically the cameras could simply send all observed data to the central office but this would be a privacy violation. Instead, the central office is an external decider and all cameras deliver input for various PSOs.
\item \textbf{Example-4: }A decentralized social networking platform such as HELIOS\footnote{HELIOS project homepage. (2021). Retrieved from https://helios-h2020.eu/.} is inherently more privacy-friendly than centralized solutions. However, people would still like to make searches in larger setting than within their own direct contacts.
For example, number of people have formed a group (for some purpose). In order to extend the group, one of them would like to identify, in privacy-preserving manner, whether her/his friend is also a friend of at least three other existing group members and could be asked to join. 
\end{itemize}
\section{Classification of PSO Problems}
The scenarios of Section 2 lead to different types of PSO problems. In general, PSO problems can be classified using different criteria. In the following, we discuss some of these criteria. Later we study some of the many possible PSO problems further and develop solutions. Note that the list of criteria is certainly not complete and there could be more factors that guide the future research directions.

\begin{itemize}
\item \textbf{Criterion 1: "What information is wanted from the set". }At least the following three questions could be asked about the end result set: "What is the cardinality of the end result set?" "What are the elements of this set?" "Is it empty?"
\item \textbf{Criterion 2: "Who gets the outcome". }PSO problems vary also in the way the final result is learned. For instance, in some scenarios it is required that all parties learn the outcome, whereas in some cases only one party learns the result. In this paper, we focus on the case where the result of the protocol goes to an external decider.
\item \textbf{Criterion 3: "Adversary model". }Semi-honest and malicious adversarial models are two common settings in privacy-preserving protocols. 
\item \textbf{Criterion 4: "Size of the universe". }In each PSO problem, elements of the input sets belong to a priori defined set of potential elements. Hereafter we call this set \textit{the universe}. Whether the total number of elements in the universe is limited or not is a criterion that could be taken into account when defining a PSO problem
\item \textbf{Criterion 5: "Number of parties that have input". }Number of participants can vary a lot, depending on the use case for which the PSO problem is solved.
\item \textbf{Criterion 6: "What is the set operation". }We need to determine what is the set operation to be computed, defined by combining intersections, unions and complements of sets.
\item \textbf{Criterion 7: "Size of input sets". }Whether the sets are the same (or almost
the same) in size, whether the sets are large or small, or whether some sets are actually singletons.
\item \textbf{Criterion 8: "Can we use a trusted third party". }
\item \textbf{Criterion 9: "On-line or off-line sets". }There are use cases where at least some part of the input data is known well before the output data is needed. Then it may be possible to do some off-line computations before the PSO protocol is run in on-line fashion. 
\end{itemize}

\section{Problem Statement}
We study many variants of multi-party PSO problems. For the criterion "Who gets the outcome" we restrict ourselves on the case where the decider gets the result. For the Criterion 8, we only cover the case without the trusted third party. For the Criterion 3, we first assume that the parties are  semi-honest. Later we show how to modify the protocols to fit in the malicious adversarial model. 

We assume there are $n \geq 2$ parties plus a decider in the protocol. Each party $P_i$ has a private set $S_i$, where $1\leq i \leq n$. Each example of Section 2 can be described as one of the following problems: 
\begin{itemize}
\item \textbf{Problem-1: }All parties have their private sets as input. After executing the protocol, the decider wants to learn answer to one of the following questions: 1) What are the elements in the union of all $n$ sets. 2) What is the cardinality of the union. 3) Whether the union is an empty set. Example 1 corresponds to this problem.
\item \textbf{Problem-2: }All parties have a set. After executing the protocol, the decider learns answer to one of the following questions: 1) What are the elements in the intersection of $n$ sets of all other parties. 2) What is the cardinality of the intersection. 3) Whether the intersection is empty. Examples 2 corresponds to this problem.
\item \textbf{Problem-3: }The parties all have their private sets. After executing the protocol, the decider learns the result of any given set operation. The general PSO can be written in Conjunctive Normal Form (CNF):
\begin{equation}
S_{T}=(A_{1,1}\cup ...\cup A_{1,\alpha_1} ) \cap ... \cap (A_{\beta,1} \cup ... \cup A_{\beta,\alpha_{\beta}}) 
\label{eq1}
\end{equation}
where $A_{i,j} \in \{S_1,...,S_n,\bar{S}_1,...,\bar{S}_n\}$, the set $\bar{S}_i$ is the complement of the set $S_i$, and $1\leq \alpha \leq n$ and $\beta \in \field{N}$.
After the protocol has been executed the decider learns answer to one of the following questions: 1) What are the elements of $S_T$. 2) What is the cardinality of $S_T$. 3) Whether $S_T$ is empty. Examples 3 and 4 require more complex set operation than straight-forward union or intersection.
\end{itemize}
Note that Problem-3 covers both Problem-1 and Problem-2 but we take those separately for two reasons: 1) so many use cases are only about Problem-1 and Problem-2, and 2) our solutions for Problem-1 and Problem-2 are used when building solution for Problem-3.
\section{Related Work}
In this section, we first present the state of the art in PSI protocols, then we present some of the notable previous works on multi-party private set operation. 

As we mentioned before, private set operations are applicable to variety of use-cases, such as privacy-preserving genomic similarity \cite{dugan2016survey} and private profile matching in social media \cite{li2011findu}. Therefore, different variants of PSO problems have been studied extensively, see \cite{pinkas2018scalable}. Kolesnikov et al. in \cite{kolesnikov2017practical} proposed a new function that is called Oblivious Programmable Pseudorandom Function, and used it to design a practical multi-party PSI protocol that is secure in a malicious setting. In 2019, Ghosh and Nilges proposed a novel approach to PSI \cite{ghosh2019algebraic}, by utilizing Oblivious Linear Function Evaluation (OLE) to evaluate the intersection.

At the time of writing, the protocol of Kolesnikov et al., in \cite{kolesnikov2016efficient} is the fastest two-party PSI protocol. In \cite{kolesnikov2016efficient}, the authors proposed a variant of Oblivious Pseudorandom Function, and utilized it to achieve a light-weight PSI protocol.

Chun et al., generalized the problem of PSO by studying any PSO problem in disjunctive normal form (DNF) \cite{chun2013privacy}. Wang et al., further studied the general PSO problem in DNF for a limited universe \cite{wang2020privacy}. In this paper, we study the general problem of PSO in the setting with an external decider. To the best of knowledge, this is the first time that this problem is studied comprehensively.

\section{Protocols}
In this section, we present our privacy preserving set operation protocols. These protocols are our proposed solutions to the problems of Section 4. 

In each protocol there are there are $n+1$ participants involved with the set operations: $n$ parties have input sets and the result of the protocol goes to the decider $D$, who does not have an input set. Other $n$ parties do not get the final outcome of the protocol. We present our protocols with the assumption that the participants are semi-honest. Later in Section 9 we present a solution for malicious model as well.

In this section we assume that the universe is limited, i.e., it is possible to present it as an ordered set $U=\{a_1,..., a_u\}$. For simplicity, we assume that the decider creates this ordered set. Protocol 0 is the off-line phase used in all protocols of this section.

Let us assume that party $P_i$ has a private input set $S_i$, for $i= 1,...,n$. The decider wants to learn one of the following cases about the union of these $n$ sets: 1) What are the elements in $\bigcup_{i=1}^{n} S_i$. 2) What is the cardinality of $\bigcup_{i=1}^{n} S_i$. Our solution for these questions is by Protocol 1.

If the application area of the protocol is such that the decider only needs to know whether the union of all input sets is empty, the protocol can be simplified a lot. Instead of vector $V$ we have just one value $V$ that is initially set to enc(1). Then, each $P_i$ multiplies $V$ by enc(0) if $S_i$ is empty and replaces $V$ by enc(0) if $S_i$ is not empty. When all the parties have altered $V$, party $P_n$ sends it to $D$. The decider decrypts and if $D$ gets zero it means that the union is non-empty. 

After executing the on-line phase, only the decider learns the outcome, and other parties do not learn anything else about this protocol than  that it was run.

Let us again assume that the party $P_i$ has the set $S_i$. The decider wants to learn one of the following cases about the intersection of these $n$ sets: 1) What are the elements in $\bigcap_{i=1}^{n} S_i$. 2) What is the cardinality of $\bigcap_{i=1}^{n} S_i$. 3) Whether $\bigcap_{i=1}^{n} S_i$ is an empty set. Our solution to these questions is Protocol 2.
\SetAlgorithmName{​​​​Protocol}{}
\SetAlgoRefName{0}
\begin{algorithm}[t]
\DontPrintSemicolon 
\nonl \textbf{The off-line phase.}\\
The decider sends the ordering of  $U=\{a_1,..., a_u\}$ to all the other parties.\\
The decider creates public and private keys that fulfil requirements for an additively homomorphic encryption scheme.\\
The decider picks one of the parties randomly and informs this party that they are responsible for sending the final result to the decider. For simplicity, let us assume that the decider chooses party $P_n$.\\
The decider sends the public key of the encryption scheme to all parties.\\
Each party $P_i$ creates two sets of encrypted values by utilizing the decider's public key. One set contains $u$ instances of enc(0), and the other set contains $u$ instances of enc($r$) where $r$  is a random number chosen specifically for that instance.\\
Parties create a shared repository. All parties can read and write to the repository but only two parties cannot write at the same time, to avoid conflicts. Neither the decider nor anybody else than parties $P_i$ have access to this repository.\\
Parties together create a vector $V = (V_1,...,V_u)$ with $u$ components, where each component is an instance of enc($r$), where $r$  is a random number.
\label{1}
\end{algorithm}

\SetAlgoRefName{1}
\begin{algorithm}[t]
\DontPrintSemicolon 
\setcounter{AlgoLine}{-1}
\nonl \textbf{Protocol-1}\\
Protocol-0 is run.\\
Each party $P_i$, where $1 \leq i \leq n$, modifies vector $V$ as follows. If $u_j \in S_i$ then $P_i$ replaces $V_j$ by enc(0), which is one of the encryptions of zero that $P_i$ generated in Protocol-0. If $u_j \notin S_i$ then $P_i$ multiplies $V_j$ by enc(0).\\
When all parties have finished their modifications on vector $V$ then one of the following cases will take place.\\
\textbf{Case 1: }Party $P_n$ sends $V$ to the decider. The decider decrypts components of this vector, and $a_j \in \bigcup_{i=1}^{n} S_i$ if and only if $dec(V_j) = 0$. \phantom{xxxxxxxxxxxxxxxx}
\textbf{Case 2: }Party $P_n$ permutes the components of $V$ before sending them to $D$. The decider decrypts $V$. The number of zeros in the decrypted vector is equal to the cardinality of $\bigcup_{i=1}^{n} S_i$.
\label{2}
\end{algorithm}

\SetAlgoRefName{2}
\begin{algorithm}[t]
\DontPrintSemicolon 
\setcounter{AlgoLine}{-1}
\nonl \textbf{Protocol-2}\\
Protocol-0 is run, with one difference: in step 7, vector $V$ is initially set to instances of enc(0).\\
Each party $P_i$ where $1 \leq i \leq n$ modifies every component $j$ of the vector $V$ as follows.
\begin{equation}
  V_{j} =
    \begin{cases}
     V_{j}\cdot enc(0) & \text{if $a_j \in S_i$}\\
      V_{j}\cdot enc(r)  & \text{otherwise.}
    \end{cases}       
\end{equation} 
After all the parties have modified  the vector $V$, one of the following cases will be executed.\\
\textbf{Case 1: }Party $P_n$ sends $V$ to the decider. After decrypting, for entries that do not decrypt to zero, $D$ learns they are not in the intersection. Therefore, the decider learns $\bigcap_{i=1}^{n} S_i$. \phantom{xxxxxxxxxxxxxxxxxxxxxxxxxxxxxxxxxxxxxxxxxxx}
\textbf{Case 2: }Party $P_n$ shuffles $V$ and sends it to $D$. The decider decrypts and the number of zero values in the decrypted vector is the cardinality of $\bigcap_{i=1}^{n}S_i$.
\textbf{Case 3: }To hide the true cardinality of the intersection, the parties need to create "clones" of elements in the universe. For each element that is in the intersection, there would be many zeros after decryption. The parties would also add many encryptions of non-zero numbers to hide further the number of elements in the intersection. \\

\label{3}
\end{algorithm}

After executing this protocol, only the decider learns the result. Other parties do not learn anything about each others' sets or about the result of the protocol.

It can be shown that every set that is obtained from a collection of sets by operations of intersection, union and complement can equivalently be computed by a conjunctive normal form. Thus, the general PSO problem can be written as presented in Equ. 1.

The general PSO problem can be formalized as follows: Party $P_i$ has a private input set $S_i$, for $i= 1,...,n$. The decider $D$ wants to learn one or more of the following cases about the set $S_{T}$ of Equation \ref{eq1}: 1) What elements are there in the set $S_{T}$. 2) What is the cardinality of $S_{T}$. 3) Whether $S_{T}$ is an empty set. Other parties should not learn anything about this protocol except that it is executed. Our solution for these questions is by Protocol 3.

\SetAlgoRefName{3}
\begin{algorithm}[t]
\DontPrintSemicolon
\setcounter{AlgoLine}{-1}
\nonl \textbf{Protocol-3}\\
The off-line phase of this generic protocol is as explained in Protocol-0 with only one difference: in the step 7, the parties create $\beta$ vectors $W^k$, where $1 \leq k \leq \beta$. Each vector $W^k$ is created similarly to the vector $V$ in step 7 of Protocol-0. On-line phase of Protocol-1 is used as a building block of the on-line phase of the generic solution.\\ 
In order to compute each $W^k$, parties should compute each set $(A_{k,1}\cup ...\cup A_{k,\alpha_k})$ by utilizing Protocol-1.  Note that party $P_i$ either inputs $S_i$ or $\bar{S}_i$ or does not attend the computation for this term.\\
After all the vectors $W^k$ have been computed, party $P_n$ creates a new vector $Z$ where every entry $j$ of the vector is computed as $Z_j=\prod_{k=1}^{\beta} W^k_j \cdot$\\
Now, one of the following cases will be executed.\\
\textbf{Case 1: }Party $P_n$ sends vector $Z$ to $D$. 
The decider decrypts $Z$. For every entry $Z_j$
which decrypts to zero, the decider learns that the corresponding element $a_j$ is in the set $S_T$. \phantom{xxxxxxxxxxxxxxxxxxxxxxxxxxxxxxxxxxxxxxxxx}
\textbf{Case 2: }Party $P_n$ shuffles vector $Z$ and sends it to $D$. The decider then decrypts the vector $Z$. The cardinality of $S_{T}$ is equal to the number of zero values in the decrypted vector.\phantom{xxxxxxxxxxxxxxxxxxxxxxxxxxxxxxxxxxxxxx}
\textbf{Case 3: }Similarly to Case 3 of Protocol-2, the party $P_n$ creates a new vector $Z'$ from the vector $Z$, by appending the components of $Z$ and their duplicates to vector $Z'$. Then, $P_n$ shuffles vector $Z'$, and sends it to the decider. 
The decider decrypts the vector. If at least one of the values in $Z'$ decrypts to zero, then $S_T$ is non-empty. Otherwise, the set of Equ. \ref{eq1} is empty.\\
\label{4}
\end{algorithm}

Next we drop the assumption that the universe is limited, and present protocols for finding answers to the following questions about the set $S_T$ that can also be written in Disjunctive Normal Form as $S_T = (A_{1,1}\cap ...\cap A_{1,\alpha_1} ) \cup ... \cup (A_{\beta,1} \cap ... \cap A_{\beta,\alpha_{\beta}})$:
1) What is the cardinality of $S_T$? 2) Is this set empty? In other words, our protocols cover all private set operations but the decider cannot get the elements in the result set. In this protocol we use keyed hash function, and assume that the parties are semi-honest.

In the off-line phase of this protocol, the parties $P_1$ to $P_n$ decide on a key $k$ to be used for computing a keyed hash function. The decider should not learn this key. 

The basic idea is  simple. All parties replace elements in their input sets with images of the elements under keyed hash function. However, the parties cannot simply send the images to the decider because then the decider would get lots of information about the input sets. 

Let us first consider the question 1. The true cardinalities of all input sets are hidden from the decider and from other parties by adding a big number of "dummies" among the true images of elements in input sets. These dummies are just random bit strings that look like results of the keyed hash function. 

We present the protocol in the case of a simple example, for better illustration. It is straight-forward to convert the presentation to the general case but we skip it, for sake of compactness. Let us assume that there are three parties $A$, $B$ and $C$ in the protocol and the decider wants to know the cardinality of set $(A\cap B \cap \bar{C}) \cup (B \cap C)$. The Venn diagram for the three input sets is shown in Figure 1. Eight disjoint sets are formed by first choosing either the input set or its complement for each party and taking an intersection of the chosen three sets. Because the total universe could be very large, one of these eight sets (the intersection of complements) is assumed to be non-relevant for the end result of the PSO. For each of the other 7 disjoint sets, the parties $A$, $B$ and $C$ agree on the number of dummy values. These numbers should be at least one order of magnitude greater than the typical size of an input set. The agreed numbers of dummy values are shown in Fig. 1. 

The parties also need to agree on the actual 97 dummy values that every party adds to the set of values they would later send to the decider. Similarly, $A$ and $C$ have to agree values for the 12 dummies that both of them include among keyed hash images of their input sets. Parties $A$ and $B$ agree on 23 joint dummy values, while $B$ and $C$ agree on 53 joint dummies. Finally, $A$ would choose 34 random dummy values while $B$ (resp., $C$) choose 88 (resp., 145) dummies.  The decider receives all the hash-values and dummy values. From the received values, the decider identifies those that appear in every set, those that have been received from $A$ and $B$ but not from $C$, and those that have been received from $B$ and $C$ but not from $A$. Finally the decider is asked to subtract 173 ( = 23 + 97 + 53 ) from the gross number of collisions explained above. 

Note that every PSO problem can be presented in CNF or alternatively in DNF. Our example PSO above was in disjunctive normal form. This is mainly just for making our solution easier to explain.

\begin{figure}[t]
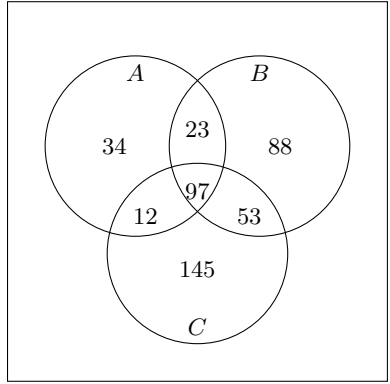

\center
\begin{venndiagram3sets}[ labelOnlyA={34},labelOnlyB={88},labelOnlyC={145},
labelOnlyAB={23},labelOnlyAC={12},labelOnlyBC={53},labelABC={97}]
\end{venndiagram3sets}
\caption{An example of a PSO for three sets with keyed hash function.\label{fig1}}
\end{figure}

Now, let us now assume that the decider only needs to learn whether the resulting set from the PSO is empty or not. In the off-line phase, each possible collection of parties would agree on several keys that would be used to compute images of elements in sets of parties in the collection. This would be done in addition to the images computed by the common key $k$. Effectively, for each element that can be found in several input sets there would be many collisions in the data received by the decider. Dummies would be added in addition to this "cloning" of elements. Apart from images by the common key $k$, parties should skip computing images of a few elements with the other keys. This is to further confuse the true number of collisions from the number of observed collisions.

\section{Performance Evaluation}
In this section we evaluate the communication and computation complexities of our protocols. We implement our protocols, and we present the results of our experiments. Moreover, we compare the performance of our protocols with the state of the art.

Please note that any set operation can be considered a boolean function, therefore, the number of possible set operations between $n$ different sets is equal to the number of truth tables with $n$ different variables, which is $2^{2^n}$. In other words, when the number of input sets increases, the number of possible set operations increases more than exponentially. This means in order to evaluate the performance of Protocol 3 and protocol with keyed hash function, we cannot run the experiment with all the possible set operations. Therefore, we choose one operation that we find interesting, and is compatible with the experiments in \cite{wang2020privacy}: In equation \ref{eq1}, we assume that $\alpha = \beta = n$.

We have compared the complexity of our Protocol 3 against the protocol 3 by Wang et al. in \cite{wang2020privacy}, because they already compared their protocol against the state of the art and showed that their protocol 3, is more feasible in practice than other solutions. It is important to note that our setting with an external decider differs from the settings in \cite{wang2020privacy} and other prior art. The difference in setting gives an opportunity to get more efficient solutions. 

\begin{table}[t]
	\centering
	\caption{Number of operations of our protocol 3 and the protocol 3 of \cite{wang2020privacy}}
	\begin{tabular}{|c c c c c c c|} 
		\hline
	Operations&&&Protocol of Wang et al. in \cite{wang2020privacy}&&& Our Protocol-3\\  [1ex]
		\hline\hline
		Encryptions&&&$O( nu+\alpha\beta u)$&&&$O(\alpha\beta u)$\\  [1ex]
		\hline
		Decryptions&&&$O(nu)$&&&$O(u)$\\  [1ex]
		\hline
	\end{tabular}
	\label{table1}
\end{table} 

Wang et al.'s protocol is based on threshold ElGamal while our protocols use Paillier encryption scheme or any additively homomorphic encryption. We have listed the number of operations in Table \ref{table1}. The entry for number of encryptions includes the number of re-encryptions. Both in ElGamal and in Paillier this is done by multiplying the encrypted value by a random encryption of zero. The speed of multiplication is very fast compared to encryption in both cryptosystems. Thus we have not listed the number of multiplications in the table. The number of operations in a single union or intersection can be calculated from the values of the table by substituting $\beta=1$.

We also compare the number of communication rounds in our protocol against the protocol of Wang et al. In \cite{wang2020privacy}, the $n$ parties first need together decide on the key for threshold ElGamal. Then each party needs individually take part in calculating the union of all sets. Then each party needs to individually take part in calculating the result and finally all parties together decrypt the result.

In our protocol the decider picks the key, then all other parties individually, in any order and maybe even partially simultaneously, modify the values in vectors $W^k$. Party $P_n$ does the multiplications and sends $Z$ to $D$ that will decrypt it. Thus the number of communication rounds is much smaller in our solution.

The time measurements are obtained by running our protocols
on an x86-64 Intel Core i5 processor clocked at 2.7 GHz with a 4 MB L3 cache. In Protocols 1, 2, and 3, any additively homomorphic non-deterministic encryption can be used. For our experiments, we use Paillier cryptosystem. The modulus that Wang et al. used in ElGamal was set to 512 bits, therefore, to have a fair comparison we set the modulus $N$ for Paillier cryptosystem to 512 bits as well.

We now compare the execution time of our Protocol 3 and the protocol 3 by Wang et al. in  \cite{wang2020privacy}. We first tested the cases where $u\in\{10, 20, 40, 60, 80, 100\}$ and there are 3 parties with input sets in the protocol. The result of our implementation showed that our protocol is 5 times faster than the protocol of Wang et al. For example, when $u=100$, our protocol 3 needs 0.71 seconds  to compute the outcome set, while protocol 3 of Wang et al. needs 3.7 seconds. We next compared the execution time of our Protocol 3 and the protocol 3 of Wang et al. in the cases that $n\in\{ 3, 5, 10, 15, 20\}$ and $u=20$. In our experiments our Protocol 3 performs significantly faster than the protocol of Wang et al., and the speed difference increases with the number of participants. For example, when $n=20$, our protocol needs 0.15 seconds, the protocol of Wang et al. needs 24 seconds to perform PSO. Therefore, when there are 20 parties with input sets, our protocol performs more that 150 times faster than Wang et al.'s protocol. Please note that we utilized the same computational power as Wang et al. used.

The evaluation of the keyed hash function is many orders of magnitude faster compare to public key encryptions. For instance, for a set size of one million, the computation of keyed hash values only takes one second.

\section{Security and Privacy Analysis}
In this section, we present the security and privacy analysis of our generic protocols in the semi-honest setting. 

In our protocols we assume that the parties communicate through a secure channel. Moreover, we also assume that the repository that the parties $P_1, ..., P_n$ use is accessible by them only. Moreover, this repository has a secure version control system to log the activities of its users \cite{guthrie2006distributed}.

In protocols with homomorphic encryption, if we assume that all parties are semi-honest, all vectors $W^k$ are calculated correctly and the correct vector $Z$ is sent to $D$. Also the result $D$ gets after decrypting $Z$ is the correct answer.

All values in the vectors in the repository are (very likely to be) different and no party $P_i$ can decrypt them without the help of $D$ and thus parties $P_i$ do not learn anything from them.

The Decider does not know the values in the repository and thus does not get any information from the parties in addition to the encrypted vector sent to it in the end of each protocol. Therefore, $D$ does not learn anything else than the decrypted values of the vector that $P_n$ send to $D$.

Thus in the semi-honest setting no party $P_i$ learns anything and the decider only learns the end result of the protocol.

In the protocols with keyed hash function, $D$ only receives the hash values of the items and dummy values. Because of the one-way property of the hash functions, $D$ cannot guess the items from their hash values. $D$ does not know the key and thus cannot even check if some element is in the resulting set or in the input set of some party. Also, hash values are indistinguishable from random dummy values, hence $D$ cannot tell these two apart. Moreover, the dummy values hide the cardinality of the elements that are not in the output set. Addition of "cloned" hash images (for the emptiness protocol) hide the true cardinality of the output set and just reveals whether the cardinality is zero or positive.

\section{Modified Protocol with One Malicious Party}
We consider now the adversarial model in which we assume that there are no collusions between the parties. If we do not assume that the parties $P_i$ are semi-honest, there are several ways how they can try to cheat: (i) They can use different inputs $S_i$ (or $\bar{S}_i$) in different parts of the protocol (in different unions); (ii) They can use an incorrect complement for their set $S_i$; (iii) They can calculate the elements in $W^k$ incorrectly: for instance, instead of multiplying  the previous value by enc(0) they replace the element by enc(r); (iv) Party $P_n$ can send an incorrect vector $Z$ to the decider.

Our protocol can be made secure against these actions in the following way:

1) We assume that $P_1$ goes first and initializes the vector $W^k$ in the same way for each $k$ such that the input set $S_1$ is included in the union. Also, $P_1$ initializes the vector $W^k$ in the same way for each $k$ such that the complement of $S_1$ is included in the union.
For each $a_i$, the initialization already includes handling of inclusion in $S_1$ (or in $\bar{S_1}$). This arrangement prevents $P_1$ from using different inputs $S_1$ (or $\bar{S_1}$) in different unions.
 
In addition, when comparing the initializations of $W^k$ for $S_i$ and that for its complement, we notice that, for each component, one of them is encryption of 0 and one is encryption of 1. Each such pair can now be sent, in random order, to the decider who can confirm that, indeed, the pair is of the form $\{enc(0, enc(1)\}$. The party $P_i$ then, for each pair, chooses a random number $a$, publishes the number, and raises both encrypted values to power $a$. The result is an (unordered) pair of enc(0) and enc($a$) for some $a$, for each element in the vector $V$. All other parties can later make sure that these are calculated correctly.

2) We repeat our protocol $n$ times and change the order of parties $P_i$ such that everyone must be once in the role of $P_1$. 

3) In every repetition every party calculates the vector $Z$ and sends it to the Decider. If every party has acted honestly in every repetition, then (i) everybody should send the same $Z$ in every repetition; (ii) the decryption of $Z$ would give the same result for each repetition. On the other hand, if some party has used different $S_i$ in some repetition than what they used when they were playing the role of $P_1$, then there is a fair chance that the results from these two repetitions do not match. 

4) The previous steps are likely to reveal whether an individual party has been cheating while others have stayed honest. However, it is harder to identify the cheater. The following can be done in order to locate where the cheating has happened. After $n$ runs of the protocol every party independently multiplies together all elements in every vector $W^k$ in every run that they know must be encryptions of zero if computations have been done correctly. Then they further multiply the result with a random encryption of zero before sending it to the Decider. The Decider decrypts everything and verifies that nothing else than zeros come out.

\section{Conclusion}
Private set operations (PSO) such as private set intersection and private set union can be used in many privacy sensitive use-cases, and therefore they have been studied extensively. In this work, we studied a special variant of PSO where the outcome of the multi-party PSO is learned by an external decider while the parties who have provided an input set to the PSO do not learn the outcome.

We presented two generic solutions (Protocol 3 and protocol with keyed hash) to any PSO problems that the decider seeks the cardinality of the output set or wants to investigate whether this set is empty. Moreover, Protocol 3 is a generic protocol for the decider to learn also elements in the output set, under the assumption that the universe, from where parties choose elements to their sets, is limited in size. In addition to the cardinality and emptiness, the decider can receive the elements in the output set. 

We presented the security and privacy analysis of our protocols in the semi-honest setting, and presented a modified protocol for the malicious setting. Lastly, we implemented our protocol. The result of our experiments showed that our solutions for our special setting, i.e., having an external decider, are more efficient than applying state of the art solutions proposed for other settings also for our setting. The experiments also show that our solutions are feasible for many real-life use cases.  
\section*{ACKNOWLEDGEMENT}

This paper is supported by 5GFORCE project funded by Business Finland, and HELIOS H2020 project. HELIOS has received funding from the European Union’s Horizon 2020 research and innovation programme under grant agreement No 825585.

%
\bibliographystyle{unsrt}

\bibliography{ref}
\end{document}